# Plasmon-enhanced second harmonic sensing

Lavinia Ghirardini[1], Anne-Laure Baudrion[2], Marco Monticelli[1], Daniela Petti[1], Paolo Biagioni[1], Lamberto Duò[1], Giovanni Pellegrini[1], Pierre-Michel Adam[2], Marco Finazzi[1], Michele Celebrano[1]

[1] *Dipartimento di Fisica, Politecnico di Milano, piazza Leonardo da Vinci 32, 20133 Milano, Italy*

[2] *Laboratoire de Nanotechnologie et d'Instrumentation Optique, Institut Charles Delaunay, Universite' de Technologie de Troyes, UMR CNRS 6281, 12 Rue Marie-Curie, CS 42060, 10004 Troyes Cedex, France*

ABSTRACT: It has been recently suggested that the nonlinear optical processes in plasmonic nanoantennas allow for a substantial boost in the sensitivity of plasmonic sensing platforms. Here we present a sensing device based on an array of non-centrosymmetric plasmonic nanoantennas featuring enhanced second harmonic generation (SHG) integrated in a microfluidic chip. We evaluate its sensitivity both in the linear and nonlinear regime using a figure of merit (FOM $= \frac{\Delta I/I}{\Delta n}$) that accounts for the relative change in the measured intensity, *I*, against the variation of the environmental refractive index *n*. While the signal-to-noise ratio achieved in both regimes allows the detection of a minimum refractive index variation $\Delta n_{min} \sim 10^{-3}$, the platform operation in the nonlinear regime features a sensitivity (i.e. the FOM) that is at least 3 times higher than the linear one. Thanks to the surface sensitivity of plasmon-enhanced SHG, our results show that the development of such SHG sensing platforms with sensitivity performances exceeding those of their linear counterparts is within reach.

TEXT

Optical sensing techniques represent one of the most promising approaches to detect and identify chemical and biological species, thanks to their non-invasive character. In this framework, platforms based on metallic structures featuring collective oscillations of the conduction electrons (surface plasmon resonances, SPRs) have rapidly become a fundamental tool for the real-time, label-free analysis of biospecific interactions and chemical reactions [1]. These sensors allow detecting small amounts of analyte through the variation in reflectivity associated with the SPR shifts induced by the local refractive index change caused by the analyte itself [2-6].

Refractometric plasmonic biosensors have been commercially available for more than 20 years as planar platforms based on propagating surface plasmon polaritons (SPPs) [7]. More recently, platforms based on localized SPR (LSPR) in metallic nanostructures have also been developed. These platforms have the potential to overcome the SPP-based approaches, since they exhibit similar molecular sensitivity [8] while requiring less sophisticated and bulky optical equipment [9]. Moreover, the high-intensity electromagnetic fields associated to the excitation of LSPRs enable a dramatic reduction of the probing volume allowing for single-molecule detection with high temporal resolution [10, 11]. Strong field confinements are also accompanied by strong field enhancements that influence many optical processes, thus broadening the range of phenomena that can be probed [2, 7, 9, 12]. These properties, together with high-end miniaturization, multiplexing and microfluidics integration, make LSPR substrates promising candidates for high-throughput screening, point-of-care real-time diagnostics, and field applications [6].

In the last years, significant efforts have been devoted to the understanding of nonlinear optical processes in plasmonic nanostructures, giving rise to the research field of nonlinear plasmonics [13]. Since nanoscale optical integration is nowadays a strategic technological perspective, the ability of generating and manipulating nonlinear optical processes at the nanoscale may allow for the implementation of nonlinear optical sensing probes and photonic sources for the next-generation technology. Yet, to date, confinement of nonlinear optical processes beyond the diffraction limit remains a challenging task since phase-matching conditions, which ensure efficient energy transfer from the fundamental to the nonlinear wave in bulk crystals, cannot be exploited in extremely small volumes. However, key strategies based on especially engineered multi-resonant plasmonic antennas have been recently proposed to effectively compensate for this problem [14-19].

So far, the refractometric plasmonic sensing scheme has been translated to the nonlinear paradigm only using the third harmonic generation (THG) process [20], demonstrating sensing performances similar to those of linear platforms in terms of signal-to-noise ratio. THG, which stems from the nonlinear currents in the volume of metals, is indeed attained with relatively high yields in plasmonic nanoantennas. Conversely, SHG has generally a very low emission yield in plasmonic nanostructures, since it is inhibited in the volume of metals for symmetry reasons. Consequently, SHG arises mainly from the uneven currents at the metal/environment interface and can be enhanced by local imperfections and asymmetries in the local geometry of the antenna or in the environment. Therefore, despite the low emission yield, its surface character makes SHG a very promising tool for plasmonic sensing and, in the last years, have stimulated the development of various concepts for plasmon-enhanced SHG sensing [16].

In this work, we realize and characterize a first platform prototype for plasmon-enhanced SHG sensing based on dense ordered arrays of L-shaped Au nanoantennas fabricated on glass and encapsulated in a microfluidic device. We evaluate its nonlinear sensing performances and compare them with the linear ones by monitoring the variation of the SHG yield and of the sample reflectance at the pump fundamental wavelength (FW) while changing the refractive index of the environment. This is achieved by putting our sample in contact with three solutions of water and ethanol at different concentrations. These periodic arrays demonstrated an enhanced SHG thanks to the lack of inversion symmetry of each single element, a double resonance at both the FW and the second harmonic (SH) wavelength, along with a sizeable spatial overlap between the plasmonic modes associated with such resonances. In addition, they feature a high directionality in the emission pattern of the SHG that allows for efficient signal collection in a relatively narrow solid angle, hence enabling the use of low NA optical detection systems [21-26] although at the expenses of the overall SHG efficiency [21].

The investigated sample comprises two identical sets of six arrays of L-shaped gold antennas (see Figure 1a), called pads from now on for the sake of simplicity. The nanoantenna arms within each pad are 50-nm-thick and 40-nm-wide and arranged in a periodic square lattice with a gap of 100 nm between adjacent antennas (see Figure 1b). Each pad is characterized by antennas featuring the same arm-length $L$ that varies from 135 to 285 nm in steps of $30 \pm 2$ nm for each pad, to tune the LSPR at different wavelengths. The plasmonic platform design and fabrication by electron-beam lithography are detailed elsewhere [21] along with its nonlinear characterization in air environment.

The device is equipped with a microfluidic channel (100 µm wide and 30 µm high) in polydimethylsiloxane (PDMS), fabricated by soft-lithography and bonded to the glass substrate with an oxygen plasma treatment. One of the two sets of pads is encapsulated inside the microfluidic channel, whereas the PDMS sticks directly onto the second set of pads (see Figure 1a), isolating it from the fluid. The latter set of pads serves as a real-time reference for the sensing experiment, since it is provided with a constant environmental refractive index ($n = 1.4$) during the entire experiment. This allows increasing the signal-to-noise ratio of the measurement by compensating for possible slow laser power fluctuations, i.e. above the integration time per each line ( $> 1 \div 10$ sec.) in the scan. The tested solutions are volume mixtures of $(1 - x)$ pure deionized water and $x$ ethanol, with $x = 0$, 25% and 50% that flow through the channel thanks to a syringe pump and are collected at the output in a drain below the sample level. Between subsequent acquisitions, the channel is emptied and air is flown to make sure that the previous solution is completely removed and that the channel is dry before flowing the next liquid.

In the linear sensing configuration, a change in the refractive index of the environment, $\Delta n$, causes a frequency shift, $\Delta \omega$, in the LSPR of the antennas, which translates into a variation of the detected reflected intensity, $\Delta I_\omega$, at the pump frequency, $\omega$. Since the SHG yield depends quadratically on the field intensity enhancement produced by the antenna resonance, a change in $\Delta n$ is expected to induce a sizeable variation in the SHG intensity, $\Delta I_{2\omega}$. The figures of merit (FOM) that we will analyze are then $\Delta(I/I_{\text{ref}})_\omega/\Delta n$ and $\Delta(I/I_{\text{ref}})_{2\omega}/\Delta n$ for the linear and nonlinear case, respectively [27], where $I_{\text{ref}}$ represents both the reflected and SHG intensity recorded from the reference pads.

A standard home-built inverted microscope (see Figure 1c) is used to characterize and compare the linear and nonlinear emission properties of the pads. The femtosecond pulses ($\Delta\tau \sim 150$ fs) from an Er-doped fiber laser are linearly polarized (~$10^2$:1) to best excite the fundamental mode of the antennas around 1550 nm, and focused on the sample through a 0.7-NA long-working-distance air objective. The signal from the sample is then collected in epi-reflection geometry through the same objective and it is sent to the detection path through a non-polarizing beam-splitter. A dichroic mirror (DMSP1000, Thorlabs Inc.) separates the nonlinear emission from the FW coming from the device, which is reflected onto a commercial InGaAs photodiode (PDA20CS, Thorlabs Inc.) to monitor the linear reflectivity. The nonlinear emissions further filtered by a narrow bandpass filter (centered around 775 nm with a 25-nm bandwidth) to isolate the SHG peak from photoluminescence and residual FW light and is detected with a silicon single photon avalanche diode (PDM Series-C module, MPD Srl). The linear and nonlinear maps are collected by raster-scanning a piezoelectric stage (P517-3CL, Physik Instrumente GmbH), after the solutions with different refractive indices reach a static condition in the microfluidic channel.

Figure 2 shows the FW (a) and SHG (b) maps collected by exciting the sample with 800 μW average power. The reference pads are the ones on the right, as confirmed by the fact that their intensity does not significantly change through the different measurements. The set of pads exposed to the solutions is the one on the left in each panel. One can immediately recognize that, as expected, the resonance shifts to longer antennas as the environment refractive index is progressively increased (especially comparing Air with pure $H_2O$ where the step index change is about 30%). This trend can be observed both in the linear (Fig. 2a) and nonlinear (Fig. 2b) maps. Due to the well-known change in the radiation pattern of the generated light that favors emission towards the higher refractive index region [28, 29], an intensity variation can be easily observed (especially in the SHG map) when the nanoantennas are measured in the solution with respect to air. The rise in the environmental refractive index induced by flowing a water-ethanol mixture in the channel increases the emission of light towards the upper half-space – i.e. away from the collection path - thus reducing the measured signal. The absence of SHG from the stripe in the middle of the sample (clearly visible in the FW maps), used for alignment purpose, further evidences the strong SHG yield of the antenna arrays.

The shift of the resonance is more clearly visible in Fig. 3, by comparing both the FW reflectance and the SHG emission of the signal pads (panels a and c) with those of the reference pads under PDMS (panels b and d) for the different refractive indices. Each point represents the average emission of a pad during four different scans with the same solution, and the error bar is the standard deviation of this average value, hence reflecting the measurement repeatability. Each scan covers the whole sample and lasts approximately 10-20 minutes. The resonance profiles from the reference pads are almost identical for the different solutions, both in the linear (Fig. 3b) and nonlinear (Fig. 3d) measurements. A comparison between the two panels also indicates a slight red shift occurring between the FW resonance and the SHG one (i.e. higher SHG yield is displayed at shorter arm lengths with respect to the peak reflectance), as already reported in literature [30-32].

Figure 3e shows the emission power curves acquired on each pad. The quadratic behavior of the SHG signal dependence on the pump power confirms that the measured signal comes from a two-photon excitation process and that the measurements are carried out below the damage threshold of the investigated structures. It is worth mentioning that, although here we select SHG using a narrow band filter, two-photon photoluminescence from gold in the visible region is dramatically reduced when pumping at telecom wavelengths [14].

Figure 4 shows the evolution of both the normalized reflectivity (panel a) and SHG emission (panel b) from the antennas as the environmental refractive index *n* is changed. Each point represents the average intensity either reflected or emitted from each individual pad normalized by that of the refence pad under PDMS, $(I/I_{\text{ref}})_\omega$ or $(I/I_{\text{ref}})_{2\omega}$. Considering an integration time of about 10 ms per pixel in our spatial maps, about 1 min acquisition time is required for each experimental data point, which is compatible with that of previously-reported nonlinear sensing experiments [20].

By relating Figs. 4 to Fig. 3, one can readily notice that the dependence of the linear reflectivity on the plasmonic resonance of the nanoantennas as *n* changes is the one expected for standard LSPR refractometric sensing. This becomes evident by observing the variation of the normalized FW reflectivity, which is the slope of the linear fit to the data in Fig. 4, as the arm length of the nanoantennas is increased (see panel a to f): it is higher for the device featuring the shortest arm length (pad 1), it decreases when approaching the resonant pad (pad 4) and it changes sign increasing its absolute value for longer arm length (pads 5 and 6). In fact, the geometrical parameters are such that when the pad response lies on the inflection points of the resonance at the FW (see Fig. 3c), a shift in the resonance curve – caused by the refractive index variation – induces the maximum signal variation. Conversely, close to the resonance peak the intensity variation $(\Delta I/I_{\text{ref}})_\omega$ reaches a minimum. In other words, an increase in the refractive index induces a red shift in the resonance (see Fig. 3a and c), therefore the resonant condition is met progressively towards shorter arm lengths, hence determining a boost (drop) of the probe signal when the pump frequency is higher (lower) than the resonance frequency of the antennas.

It is now interesting to address the SHG sensing performances of this platform and compare them to the linear ones. To evaluate the intrinsic sensitivity of each device, we computed the FOM values by the slope of each graph both in the linear and nonlinear regime, $(\Delta I/I_{\text{ref}})_\omega/\Delta n$ and $(\Delta I/I_{\text{ref}})_{2\omega}/\Delta n$ (see Fig. 5a). Higher values of the FOM are found for pad1 and are up to about 30 and 10 in the SHG and FW regime, respectively. Therefore, in this platform, plasmon-enhanced SHG sensing reaches up to 3 times higher sensitivity with respect to the linear regime. The absolute values of the FOM reported here are comparable to and even higher than those of other highly-efficient plasmonic sensing platforms operating in the linear regime [33, 34]. It is also worth noting that, for pads closely resonant with the FW, the FOM values drop to zero in the linear regime, while in the nonlinear regime they are still considerably large and more significantly overlap to the highest SHG signal. These key features suggest that the resonance shift is not the only mechanism at play in plasmon-enhanced SHG sensing and make this nonlinear sensing platform a highly interesting and competitive approach compared to the linear counterpart.

To attain a comprehensive comparison between the linear and nonlinear performances of the platform, we also determined its detection limits in both regimes by estimating the minimum measurable variation in the refractive index, $\Delta n_{\min}$ [20]. This quantity, which is commonly referred to as "resolution" in SPR sensing [35], allows to assess the sensitivity (FOM) against the signal-to-noise ratio for each measurement. $\Delta n_{\min}$ is here obtained dividing the average between the error bars in each panel of Fig. 4 (i.e. minimum detectable intensity variation) by the slope of the linear fit in the same panel, while the $\Delta n_{\min}$ error bars (see Fig. 5b) the standard deviation between the error bars constitutes.

Analyzing pad 1 according to this parameter, we find $\Delta n_{\min}^{2\omega} = \frac{(\Delta I/I_{\text{ref}})_{\min}}{(\Delta I/I_{\text{ref}})^{2\omega}/\Delta n} \sim 1.2 \times 10^{-3}$ and $\Delta n_{\min}^{\omega} \sim 6 \times 10^{-4}$ in the nonlinear and linear regime, respectively. By evaluating $\Delta n_{\min}$ for all pads in both regimes, we find that in the linear regime the best-performing devices are pads 5 and 6, featuring $\Delta n_{\min}^{\omega} \sim 4 \times 10^{-4}$, which is 3 times lower than the $\Delta n_{\min}^{2\omega}$ of pad 1 (i.e. the best in the nonlinear regime). The same pads, being out of resonance with respect to the FW, show significant performance degradation in the nonlinear regime due to the SHG signal drop. Comparable results were obtained on different samples realized using the same nominal fabrication parameters, achieving rather small variations in these values ($\Delta n^{2\omega}_{\min} \sim 0.9 \div 2 \times 10^{-3}$ and $\Delta n^{\omega}_{\min} \sim 6 \times 10^{-4}$). Given the discretization applied to the arm lengths, we ascribe these oscillations in the determination of $\Delta n_{\min}$ to uncertainties in the actual antenna geometry for the same fabrication parameters.

In conclusion, by analyzing the performances of a plasmonic sensing platform in both the linear and nonlinear regime through the variation in the environmental refractive index, we found similar performances in terms of the minimum detectable refractive index $\Delta n_{\min}$. This result, as already highlighted in a recent paper on THG plasmonic sensing [20], seems to put the linear and nonlinear regimes on an equal footage, with the former slightly favored due to its simpler implementation. Nevertheless, plasmon-enhanced SHG sensing still holds promise to overcome plasmonic sensing schemes working in the linear regime, thanks to the nonlinear dependence on the local field enhancements. In fact, the FOM of these L-shaped nanoantenna-based platforms in the nonlinear regime (i.e. ~ 30) are significantly higher than that featured in the linear regime by the same platform and by other effective platforms [33, 34]. Yet, despite offering a good tradeoff between nanostructure simplicity and their nonlinear efficiency, these devices are not quite fully optimized for efficient SHG [14]. By combining nanoantenna geometries featuring stronger nonlinearities together with a nonlinear setup with a higher signal-to-noise ratio, we envision that nonlinear plasmonic sensing platforms will overcome standard linear platforms. This could also be favored by the physical mechanisms at work in plasmon-enhanced SHG platforms, where the local binding of a nano-object to the hot spot region in properly functionalized platforms strongly modifies the local symmetry of the system and might promote further SHG modulation.

**Acknowledgements**

The authors would like to thank Prof. R. Bertacco for fruitful discussions on the microfluidic part. As the nanofabrication process was carried out through the facilities of the Nano'Mat platform (www.nanomat.eu), the authors acknowledge the financial supports from the "Ministère de l'Enseignement Supérieur et de la Recherche", the "Conseil Régional Champagne-Ardenne", the "Fonds Européen de Développement Régional (FEDER) fund", and the "Conseil Général de l' Aube". This work has been carried out in the framework of Cost Action MP1302 Nanospectroscopy. The authors would also like to acknowledge the financial support of Cariplo Foundation through Project SHAPES (2012-0736).

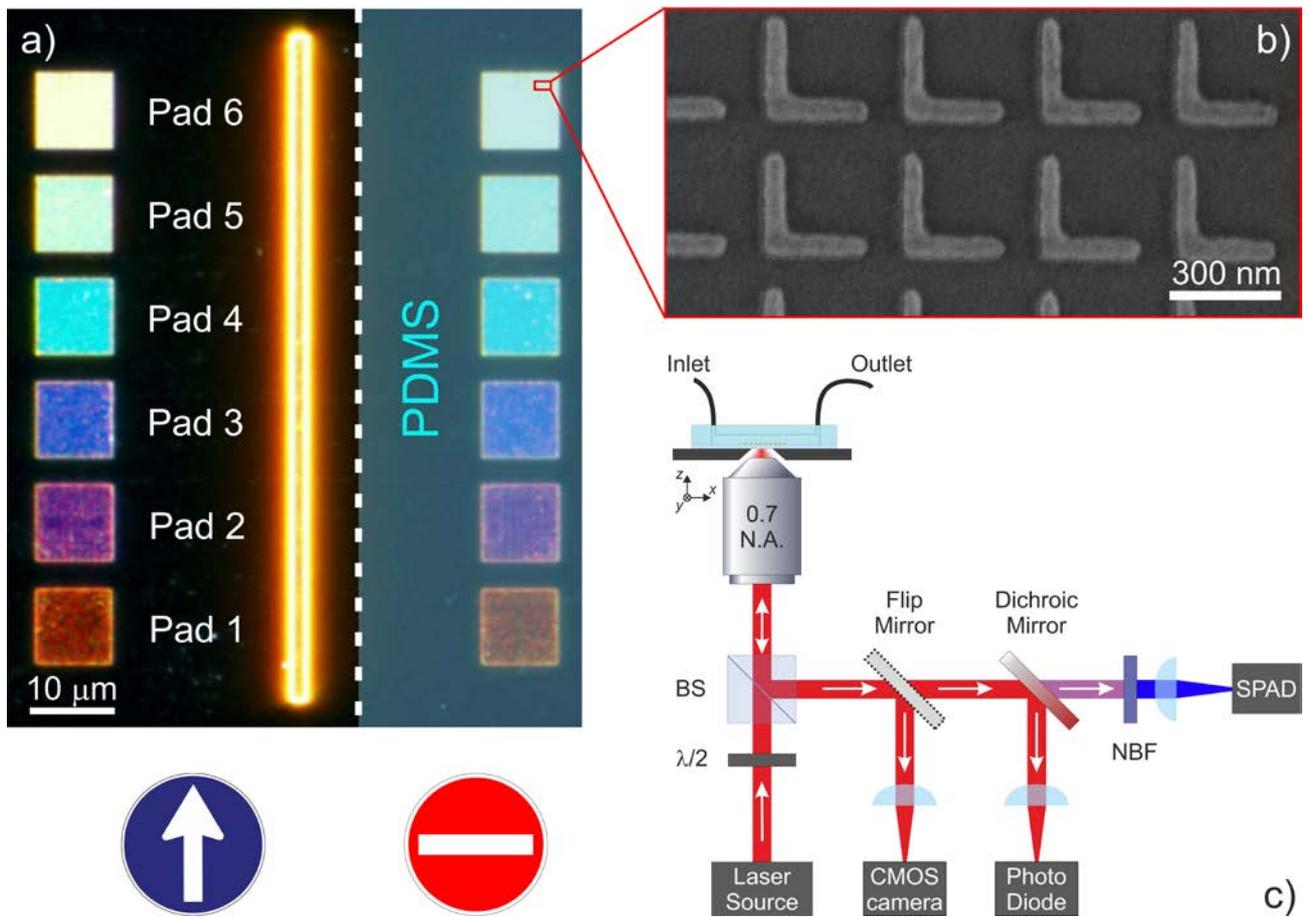

**Figure 1**: (a) Dark-field reflection map of one of the plasmonic platforms. The L-shaped antennas are arranged in two columns of six 10 µm × 10 µm square lattices on a glass substrate. The right column, covered by PDMS (transparent light-blue layer), is used as a reference to normalize the signal collected from the left column. The gold stripe in the middle is used for the microfluidic channel alignment. (b) Scaning electron microscopy map of the array with arm-length 285 nm (P6 – area maked red in panel a), recorded at a very low acceleration voltage on the non-conductive substrate. (c) Experimental setup. Laser source: Er-doped fiber laser (150-fs-long pulses at 80 MHz centered at 1554 nm). BS: non-polarizing beam-splitter. Dichroic mirror: DMSP1000 (Thorlabs Inc.). PhotoDiode: infrared InGaAs photodiode (Thorlabs Inc.). NBF: narrow band pass filter (775 nm / 25 nm badwidth). SPAD: single photon avalanche diode (PDM Series-C module, MPD S.r.l.).

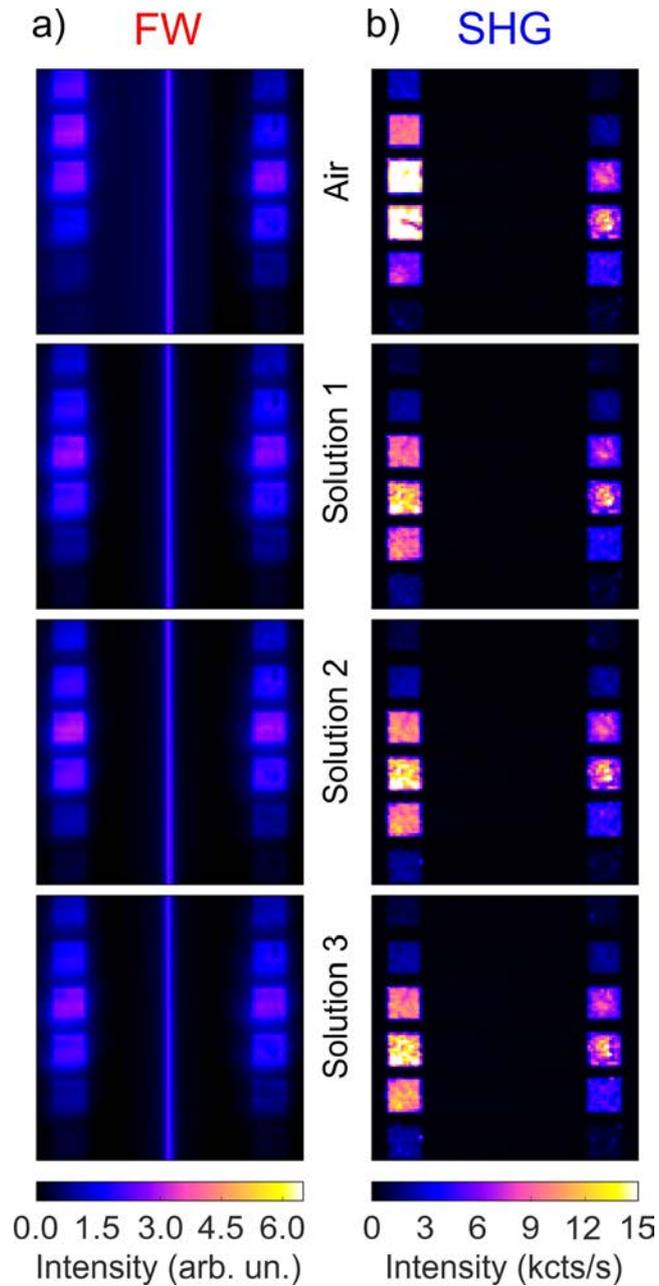

**Figure 2**: Linear (a) and nonlinear (b) maps recorded with the setup described in figure 1.b. The higher SHG yield is recored on pad 3 (arm length L = 195 nm), in agreement with what reported in Ref. 21. The array of pads on the right lies under the PDMS and remains in contact with a constant refractive index trhroughout the entire experiment. The pads on the left of the dashed line (see upper-left panel and Figure 1a) are located inside the microfluidic channel and the environmental refractive index they experience is changed during the different measurements. The tested solutions are volume mixtures of $(1 - x)$ pure deionized (DI) water and $x$ ethanol (Solution 1: $x = 0\%$, Solution 2: $x = 25\%$, Solution 3: $x = 50\%$).

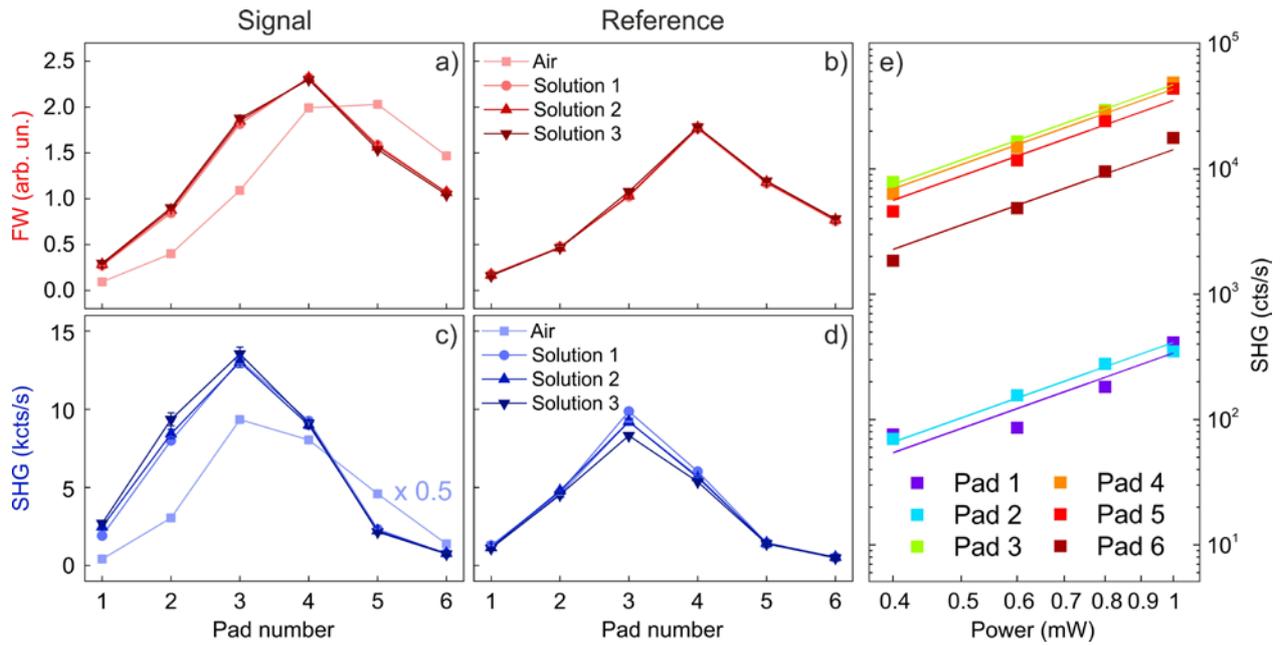

**Figure 3**: Evolution of the pads emission at the SHG wavelength (a, b) and of their reflectance at the fundamental wavelength (c, d) from both the signal and the reference pads. The tested solutions are volume mixtures of $(1 - x)$ pure deionized (DI) water and $x$ ethanol (Sol 1: $x = 0\%$, Sol 2: $x = 25\%$, Sol 3: $x = 50\%$). (e) Power curves for the pads in air. The slope of the linear fit is reported for each pad in parentesis.

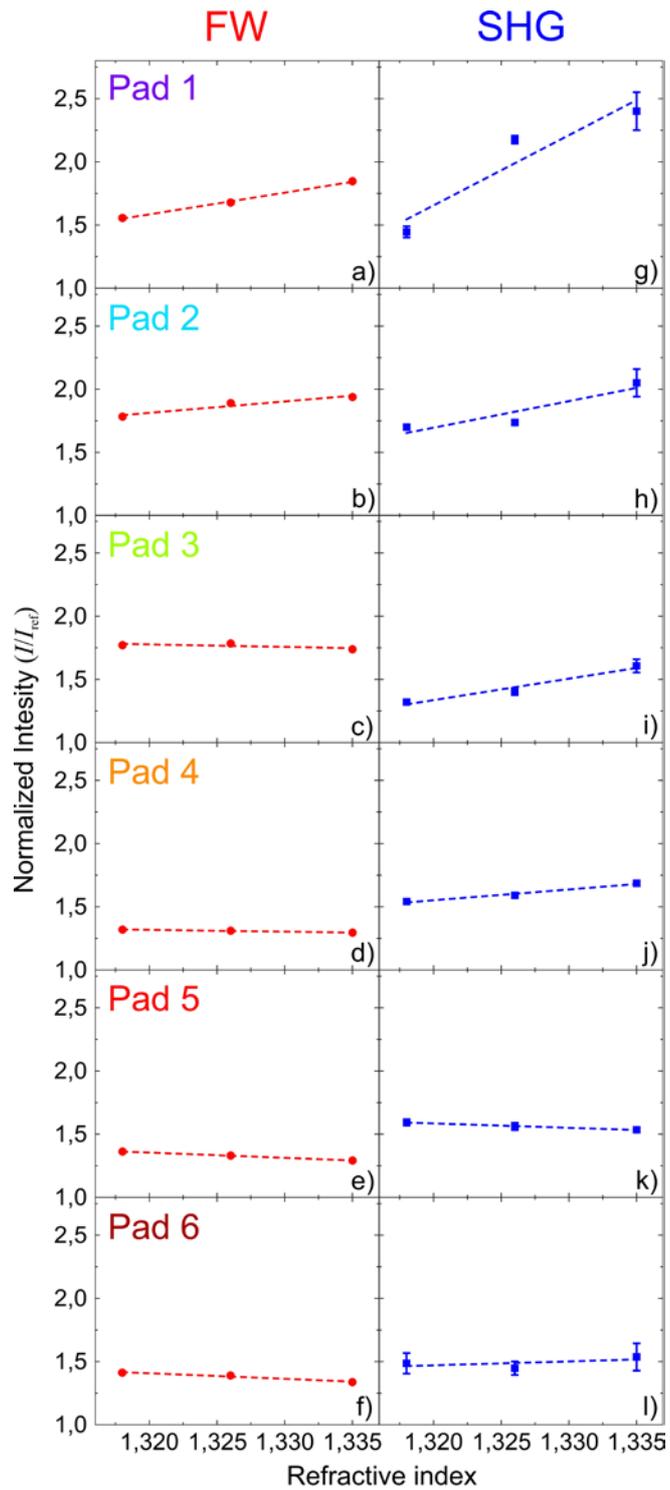

**Figure 4**: Evolution of the normalized FW reflectivity (left column, red dots) and of the normalized SHG (right column, blue squares) recorded from the six pads as a function of the environmental refractive index for the three different solutions.

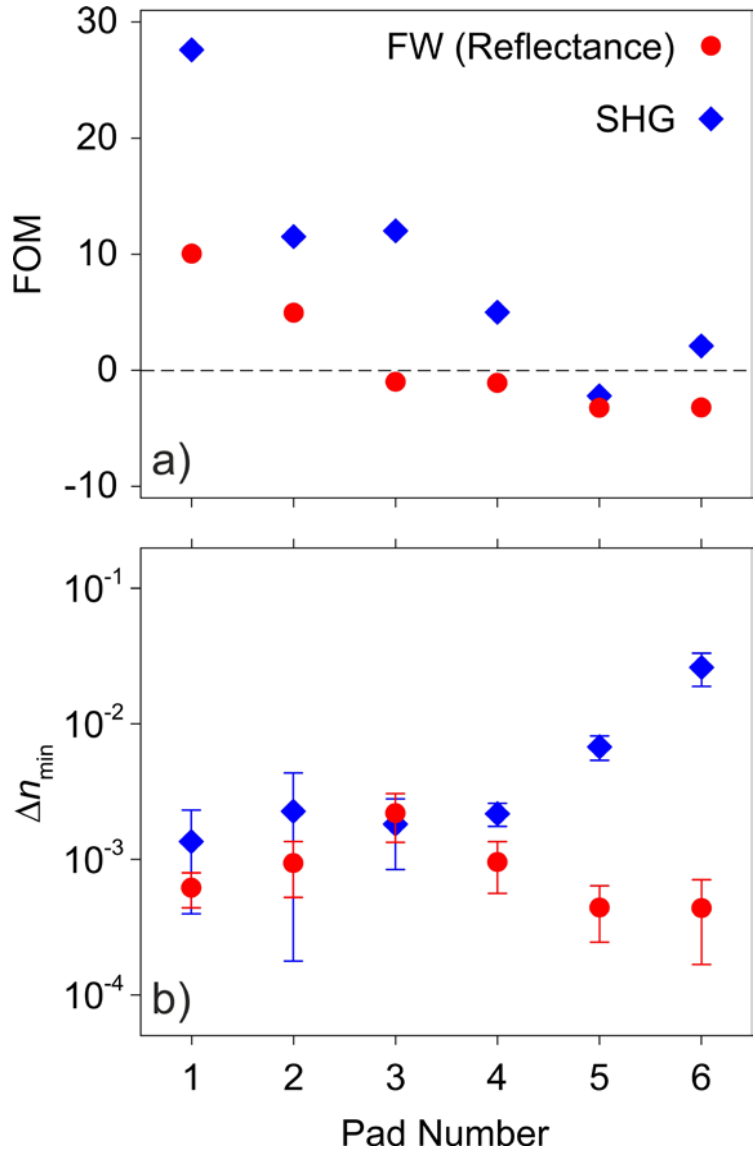

**Figure 5**: a) FOM (i.e. sensitivity) of each pad that composes our sensing platform, evaluated in both the linear (red dots) and nonlinear (blue diamonds) regime, respectively. The error bars, which represent the error associated with the linear fit of each panel in Figure 4, are contained within the symbol size. b) Minimum detectable refractive index variation, $\Delta n_{min}$, (i.e. resolution) by each pad in our sensing platform plotted in Log scale and obtained in both the linear (red dots) and nonlinear (blue diamonds) regime, respectively. The error bars correspond to the standard deviation associated to the averaging process between the three experimental uncertainties in each panel of Figure 4.

**References**

[1] Maier, S. A. Plasmonics: Fundamentals and Applications. *Springer* **2007**.

[2] Li, M., Cushing, S. K., Wu, N. Plasmon-enhanced optical sensors: a review. *Analyst* **2015**, 140, 386-406.

[3] Zhao, Y., Cao, L., Ouyang, J., Wang, M., Wang, K., Xia, X.-H. Reversible Plasmonic Probe Sensitive for pH in Micro/Nanospaces Based on i-Motif-Modulated Morpholino-Gold Nanoparticle Assembly. *Analytical Chemistry* **2013**, 85, 1053–1057.

[4] Virk, M., Xiong, K., Svedendahl, M., Käll, M., Dahlin, A. B. A Thermal Plasmonic Sensor Platform: Resistive Heating of Nanohole Arrays. *Nano Letters* **2014**, 14, 3544–3549.

[5] Langhammer, C., Zorić, I., Kasemo, B. Hydrogen Storage in Pd Nanodisks Characterized with a Novel Nanoplasmonic Sensing Scheme. *Nano Letters* **2007**, **7**, 3122–3127.

[6] Acimovic, S., Ortega, M. A., Sanz, V., Berthelot, J., Garcia-Cordero, J. L., Renger, J., Maerkl, S. J., Kreuzer, M. P., Quidant, R. LSPR Chip for Parallel, Rapid, and Sensitive Detection of Cancer Markers in Serum. *Nano Letters* **2014**, 14, 2636–2641.

[7] Editorial. Commercializing Plasmonics. Nature Photonics 2015, **9**, 477.

[8] Svedendahl. M., Chen, S., Dmitriev, A., Käll, M. Refractometric Sensing Using Propagating versus Localized Surface Plasmons: A Direct Comparison. *Nano Letters* **2009**, 9), 4428–4433.

[9] Mayer, K. M., Hafner, J. H. Localized Surface Plasmon Resonance Sensors. *Chemical Reviews* **2011**, 111, 3828–3857.

[10] Rosman, C., Prasad, J., Neiser, A., Henkel, A., Edgar, J., Sönnichsen, C. Multiplexed Plasmon Sensor for Rapid Label-Free Analyte Detection. *Nano Letters* **2013**, 13, 3243–3247.

[11] Ament I., Prasad, J., Henkel, A., Schmachtel, S., Sönnichsen, C. Single Unlabeled Protein Detection on Individual Plasmonic Nanoparticles, *Nano Letters* **2012**, 12, 1092–1095.

[12] Anker, J. N., Hall, W. P., Lyandres, O., Shah, N. C., Zha, J., Van Duyne, R. P. Biosensing with plasmonic nanosensors. *Nature Materials* **2008**, **7**, 442–453.

[13] Kauranen, M., Zayats, A. V. Nonlinear plasmonics. *Nature Photonics* **2012**, 6, 737–748.

[14] Celebrano M., Wu, X., Baselli, M., Großmann, S., Biagioni, P., Locatelli, A., De Angelis, C., Cerullo, G., Osellame, R., Hecht, B., Duò, L., Ciccacci, F., Finazzi, M. Mode matching in multiresonant plasmonic nanoantennas for enhanced second harmonic generation. *Nature Nanotechnology* **2015**, 10, 412–417.

[15] Metzger, B., Hentschel, M., Schumacher, T., Lippitz, M., Ye, X., Murray, C. B., Knabe, B., Buse, K., Giessen, H. Doubling the efficiency of third harmonic generation by positioning ITO nanocrystals into the hot-spot of plasmonic gap-antennas. *Nano Letters* **2014**, 14, 2867–2872.

[16] Butèt J., Brevet, P.-F., Martin, O. J. F. Optical Second Harmonic Generation in Plasmonic Nanostructures: From Fundamental Principles to Advanced Applications. *ACS Nano* **2015**, 9, 10545–10562.

[17] Thyagarajan, K., Rivier, S., Lovera, A., Martin, O. J. F. Enhanced Second-Harmonic Generation from Double Resonant Plasmonic Antennae. *Optics Express* **2012**, 20, 12860–12865.

[18] Aouani, H., Rahmani, M., Navarro-Cía, M., Maier, S. A. Third-harmonic upconversion enhancement from a single semiconductor nanoparticle coupled to a plasmonic antenna. *Nature Nanotechnology* **2014**, 9, 290–294.

[19] Aouani, H., Navarro-Cia, M., Rahmani, M., Sidiropoulos, T. P. H., Hong, M., Oulton, R. F., Maier, S. A. Multiresonant Broadband Optical Antennas as Efficient Tunable Nanosources of Second Harmonic Light. *Nano Letters* **2012**, 12, 4997–5002.

[20] Mesch, M., Metzger, B., Hentschel, M., Giessen, H. Nonlinear Plasmonic Sensing. *Nano Letters* **2016**, 16, 3155–3159.

[21] Baselli, M., Baudrion, A.-L., Ghirardini, L., Pellegrini, G., Sakat, E., Carletti, L., Locatelli, A., De Angelis, C., Biagioni, P., Duò, L., Finazzi, M., Adam, P.-M., Celebrano, M. Plasmon-Enhanced Second Harmonic Generation: from Individual Antennas to Extended Arrays. *Plasmonics* **2016**, 12, 1595–1600.

[22] Black L.-J., Wiecha, P. R., Wang, Y., de Groot, C. H., Paillard, V., Girard, C., Muskens, O. L., Arbouet, A. Tailoring second-harmonic generation in single L-shaped plasmonic nanoantennas from the capacitive to conductive coupling regime. *ACS Photonics* **2015**, 2, 1592–1601.

[23] Keren-Zur, S., Avayu, O., Michaeli, L., Ellenbogen, T. Nonlinear beam shaping with plasmonic metasurfaces. *ACS Photonics* **2016**, 3, 117–123.

[24] Czaplicki, R., Kiviniemi, A., Laukkanen, J., Lehtolahti, J., Kuittinen, M., Kauranen, M. Surface lattice resonances in second-harmonic generation from metasurfaces. *Optics Letters* **2016**, 41, 2684-2687.

[25] Stokes N., Cortie, M. B., Davis, T. J., McDonagh, A. M. Plasmon resonances in V-shaped gold nanostructures. *Plasmonics* **2012**, **7**, 235–243.

[26] Vercruysse, D., Sonnefraud, Y., Verellen, N., Fuchs, F. B., Di Martino, G., Lagae, L., Moshchalkov, V. V., Maier, S. A., Van Dorpe, P. Unidirectional side scattering of light of light by a single-element nanoantenna. *Nano Letters* **2013**, 13, 3843–3849.

[27] Becker, J., Trügler, A., Arpad Jakab, A., Hohenester, U., Sönnichsen, C. The Optimal Aspect Ratio of Gold Nanorods for Plasmonic Bio-sensing. *Plasmonics* **2010**, 5, 161–167.

[28] Lukosz W., Kunz, R. E. Light emission by magnetic and electric dipoles close to a plane interface. I. Total radiated power. *Journal of the Optical Society of America* **1977**, 67, 1607-1615.

[29] Chung. K., Tomljenovic-Hanic, S. Emission Properties of Fluorescent Nanoparticles Determined by Their Optical Environment. *Nanomaterials*, **2015**, 5, 895–905.

[30] Hubert, C.; Billot, L.; Adam, P.-M.; Bachelot, R.; Royer, P.; Grand, J.; Ginder, D.; Dorkenoo, K. D.; Fort, A. Role of surface plasmon in second harmonic generation from gold nanorods. *Appl. Phys. Lett.*, **2007**, 90, 181105.

[31] Metzger, B., Gui, L., Fuchs, J., Floess, D., Hentschel, M., Giessen, H. Strong Enhancement of Second Harmonic Emission by Plasmonic Resonances at the Second Harmonic Wavelength. *Nano Letters*, **2015**, 15, 3917–3922.

[32] M. Celebrano, M. Savoini, P. Biagioni, M. Zavelani-Rossi, P.-M. Adam, L. Duò, G. Cerullo, M. Finazzi, Retrieving the Complex Polarizability of Single Plasmonic Nanoresonators, Phys. Rev. B, **2009**, 80, 153407.

[33] Cattoni, A, Ghenuche, P., Haghiri-Gosnet, A.-M., Decanini, D., Chen, J., Pelouard, J.-L., Collin, S. $\lambda^3$/1000 plasmonic nanocavities for biosensing fabricated by soft UV nanoimprint lithography. *Nano Letters* **2011**, 11, 3557–3563.

[34] Lodewijks, K., Van Roy, W., Borghs, G., Lagae, L., Van Dorpe, P. Boosting the Figure-Of-Merit of LSPR-Based Refractive Index Sensing by Phase-Sensitive Measurements. *Nano Letters* **2012**, 12 (3), 1655–1659.

[35] Homola, J., Surface Plasmon Resonance Sensors for Detection of Chemical and Biological Species. *Chem. Rev.* **2008**, 108, 462–493.